\documentclass[reprint,amsmath,amssymb,aps,prl,superscriptaddress]{revtex4-1}

\usepackage{graphicx}% Include figure files
\usepackage{dcolumn}% Align table columns on decimal point
\usepackage{bm}% bold math
\usepackage[dvipsnames]{xcolor}
\usepackage{soul}
\usepackage{siunitx}

\newcommand{\gdot}{\dot{\gamma}}

\newcommand{\appropto}{\mathrel{\vcenter{\offinterlineskip\halign{\hfil$##$\cr
    \propto\cr\noalign{\kern2pt}\sim\cr\noalign{\kern-2pt}}}}}

\begin{document}

\preprint{APS/123-QED}

\title{Dynamic vorticity banding in discontinuously shear thickening suspensions}

\author{R. N. Chacko}
 \affiliation{Department of Physics, Durham University, Science Laboratories,
  South Road, Durham DH1 3LE, United Kingdom}
\author{R. Mari}
 \affiliation{Univ. Grenoble Alpes, CNRS, LIPhy, 38000 Grenoble, France}
\author{M. E. Cates}
 \affiliation{DAMTP, Centre for Mathematical Sciences, University of Cambridge,
Wilberforce Road, Cambridge CB3 0WA, United Kingdom}
\author{S. M. Fielding}
 \affiliation{Department of Physics, Durham University, Science Laboratories,
  South Road, Durham DH1 3LE, United Kingdom}

\date{13 June 2018}

\begin{abstract}

It has recently been argued
that steady-state vorticity bands cannot arise in shear
thickening suspensions, because the normal stress imbalance across the
interface between the bands will set up particle migrations. In this
Letter, we develop a simple continuum model that couples shear thickening to particle
migration. We show by linear stability analysis that homogeneous flow is unstable towards vorticity
banding, as expected, in the regime of negative
constitutive slope. In full nonlinear computations, we show however that the resulting
vorticity bands are unsteady, with spatiotemporal patterns
governed by stress-concentration coupling.  We furthermore show that
these dynamical bands also arise in direct particle simulations, in
good agreement with the continuum model.
\end{abstract}

\maketitle

Recent years have seen rapid advances in understanding the rheology of dense
non-Brownian suspensions, comprising solid particles in a
Newtonian fluid at volume fraction $\phi$ close to isotropic
jamming.  In particular, the phenomenon of shear
thickening~\cite{M&W,Brown2014}, in which the viscosity
increases with shear stress $\sigma$, has recently been
understood as an evolution from lubricated to
frictional particle interactions, as the hydrodynamic forces that
push particles together overcome short-ranged repulsive forces
keeping them apart~\cite{Lootens2005,Bashkirtseva2009,
Brown2012,Fernandez2013, Seto2013,
Heussinger2013, Mari2014, Mari2015, Lin2015a, Guy2015,
Comtet2017,Clavaud2017,Hsiao2017,
Hsu2018}. When strong, this effect creates,
for states of homogeneous shear rate $\gdot$, a
constitutive curve $\sigma(\gdot)$ that is S-shaped~\cite{Wyart2014,Mari2015a}:
a positively sloping
hydrodynamic branch of low viscosity at low stresses connects to
a positively sloping frictional branch of high viscosity at high
stresses via a negatively sloped region at intermediate stresses. A slowly
increasing imposed shear rate then provokes a discontinuous jump,
between the low and high viscosity branches, in the measured or `macroscopic' flow curve. This is known as
discontinuous shear thickening (DST)~\cite{M&W,Brown2014}.

At imposed macroscopic shear stress, when $\mathrm{d} \sigma /\mathrm{d}
\gdot < 0$ one expects homogeneous steady flow to be unstable, at least for large system sizes~\cite{Yerushalmi1970}.
(For the system sizes used in particle-based simulations, this expectation is not always met~\cite{Mari2015a}.)
Consistent with this expectation, an S-shaped constitutive curve as
described above admits (in principle) steady states comprising layers
of material coexisting at a common shear rate but with different shear
stresses. These are force-balanced so long as they stack with normals
in the vorticity direction, and are then known as ``vorticity
bands''~\cite{Olmsted2008}.
In dense suspensions, however, steady state
vorticity bands are argued to be ruled out by
the differences in normal stress that generally arise across the
interface between bands, leading to a particle migration flux~\cite{Hermes2016}.
Suggestively, experiments on suspensions and a modelling and simulation study on
the related system of dry frictional grains have
revealed an unsteady strain rate signal under conditions of constant
imposed macroscopic shear stress in the DST regime, with complicated
time dependence~\cite{Neuville2012,Hermes2016,Bossis2017,Grob2016}.

In this Letter, we advance the understanding of dynamic vorticity banding in dense suspensions.
First, we propose a scalar continuum constitutive model for
the relevant rheology, by combining the Wyart-Cates
theory~\cite{Wyart2014} (which captures shear thickening but assumes
homogeneous flow) with a suspension balance model of particle
migration~\cite{Nott1994,Morris1999,Yurkovetsky2008,Lhuillier2009,Nott2011}.
Second, for this model we use linear
stability analysis to determine when a homogeneous
shear flow is unstable to fluctuations along the vorticity axis, finding instability whenever
$\mathrm{d}\sigma/\mathrm{d}\gdot<0$ in the limit of large system size. Third, we
elucidate numerically the model's full nonlinear vorticity-banding dynamics,
identifying two distinct spatio-temporal patterns that we shall term
``travelling bands'' (TB) and ``locally oscillating bands'' (LOB).
The LOB state shows an oscillating bulk shear rate signal, as seen
experimentally~\cite{Hermes2016,Bossis2017}.  Finally,
we perform particle-based simulations using the so-called Critical
Load Model~\cite{Mari2014} and show that this also has TB and
(at least transiently) LOB states, in close counterpart to the continuum model.

\begin{figure}
\centering
\includegraphics[width=.5\textwidth]{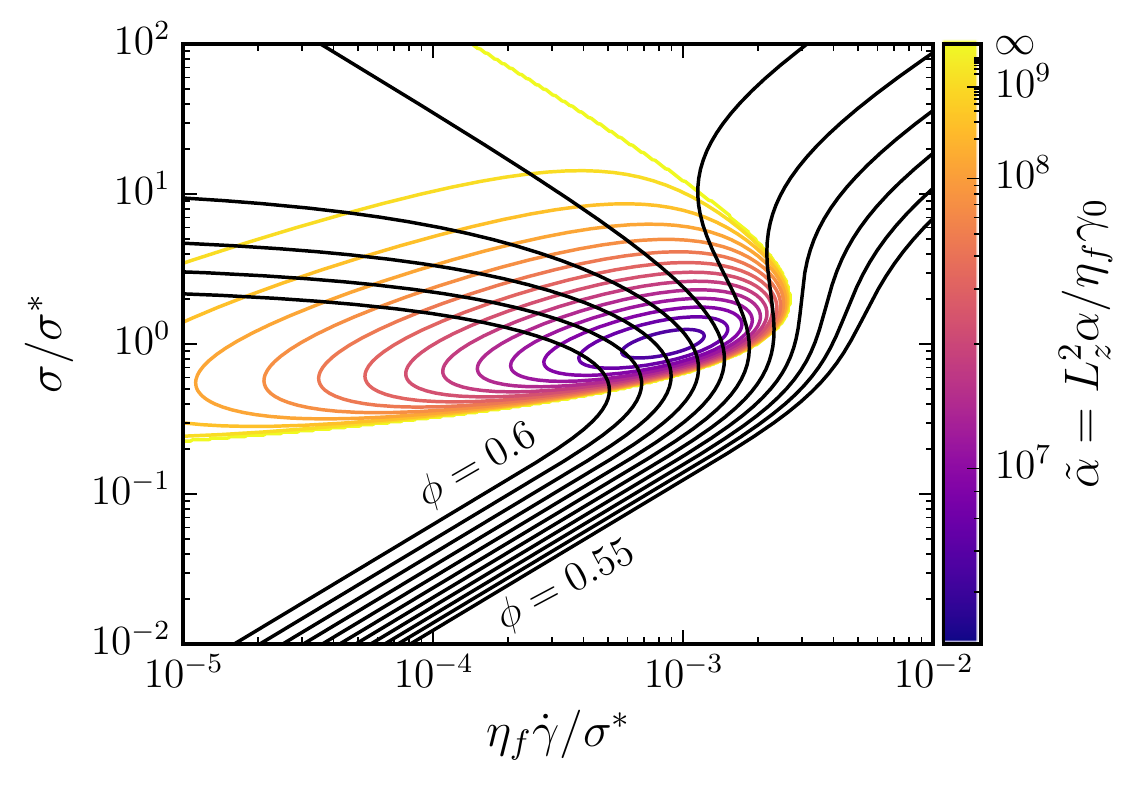}
\caption{Nondimensionalized homogeneous constitutive curves (black lines) for volume fractions $\phi \in \left[ 0.55,0.6 \right] $,
and stability boundaries (coloured lines) enclosing the region of unstable homogeneous flow
 for several $\tilde\alpha \in \left[ 5 \times 10^6, \infty \right]$, for $\tilde{\alpha}^{-1/2}$ values linearly spaced.
A given value of $\tilde{\alpha}^{-1/2}$ corresponds to a given value of the inverse system size $a/L_z$, for fixed bulk rheology parameters $\eta_f$ and $\gamma_0$.
\label{fc_fig}}
\end{figure}

We consider Stokes flow of a dense suspension sheared between hard
flat plates at $y=0,L_y$ under conditions of constant imposed shear
stress. This produces a velocity field $\bar{\bm{v}}(y) =
\left( \gdot(t) y,0,0 \right)$ with a shear rate $\gdot(t)$ that is in general time-dependent.
(Here $\bar{\bm{v}}$ denotes the suspension velocity averaged over the particle and solvent components introduced below.)
 The velocity is along the $x$ direction, its gradient along $y$, and the vorticity direction is $z$. (See \cite{SM} for a diagram.)
 As appropriate to describe vorticity banding, we assume spatial invariance in the flow direction $x$ and flow-gradient
 direction $y$, allowing spatial variations only along $z$. There can be no such variation for the shear rate $\gdot(t)$ which follows the relative
 speed of the plates.
 The dynamical variables that we consider are therefore the component
 $\sigma_{zz}(z,t)$ of the particle phase stress tensor (whose behavior is similar to that of $\sigma_{xy}(z,t)$~\cite{Wyart2014}), the fraction of frictional
 contacts $f(z,t)$, which is the microstructural order parameter entering the Wyart-Cates theory~\cite{Wyart2014},
 and the volume fraction $\phi(z,t)$.
Note that the $zz$ component of stress is actually negative in dense suspensions~\cite{M&W},
but we work throughout with its absolute value and denote this simply by $\sigma_{zz}$.
  
The Wyart-Cates  theory~\cite{Wyart2014} gives a scalar constitutive
model for the steady state homogeneous shear rheology of dense
suspensions.  While initially presented as a model for the shear stress
$\sigma_{xy}$, this equally describes $\sigma_{zz}$, because all
stress components evolve in a similar way near
jamming~\cite{Boyer2011} and across DST~\cite{Mari2014}.  The
associated viscosity is taken to diverge as the volume fraction $\phi$
approaches a critical jamming point $\phi_{\mathrm{J}}$:
\begin{equation}
\label{stress}
\eta(\phi,\phi_{\mathrm{J}}) \equiv \sigma_{zz}/\gdot=\eta_0 (\phi_{\mathrm{J}} - \phi )^{-\nu},
\end{equation}
where $\eta_0$ is, within the range of $\phi$ of interest here,
of order the solvent viscosity $\eta_{\mathrm{f}}$ and effectively constant,
so hereafter we set $\eta_0=\eta_{\mathrm{f}}$ for simplicity;
$\nu$ is likewise a constant.
Shear thickening is then captured by
assuming that at low stresses repulsive forces maintain a lubrication
film between particles, with a fraction of frictional contacts
$f\approx 0$, whereas at high stresses frictional contacts dominate
the rheology, $f\approx 1$.  The critical volume fraction for jamming
depends smoothly on stress, varying linearly with $f$ to connect 
the critical value $\phi_{\mathrm{J}}^0$ for frictionless jamming at low stresses to the
one for frictional particles $\phi_{\mathrm{J}}^\mu<\phi_{\mathrm{J}}^0$ at large stresses:
\begin{equation}
\label{jam}
\phi_{\mathrm{J}}(f) = \phi_{\mathrm{J}}^0 - f \left( \phi_{\mathrm{J}}^0 - \phi_{\mathrm{J}}^\mu \right).
\end{equation}
Here $\mu$ is the particle friction coefficient.

\begin{figure}
\centering
\includegraphics[width=.437\textwidth]{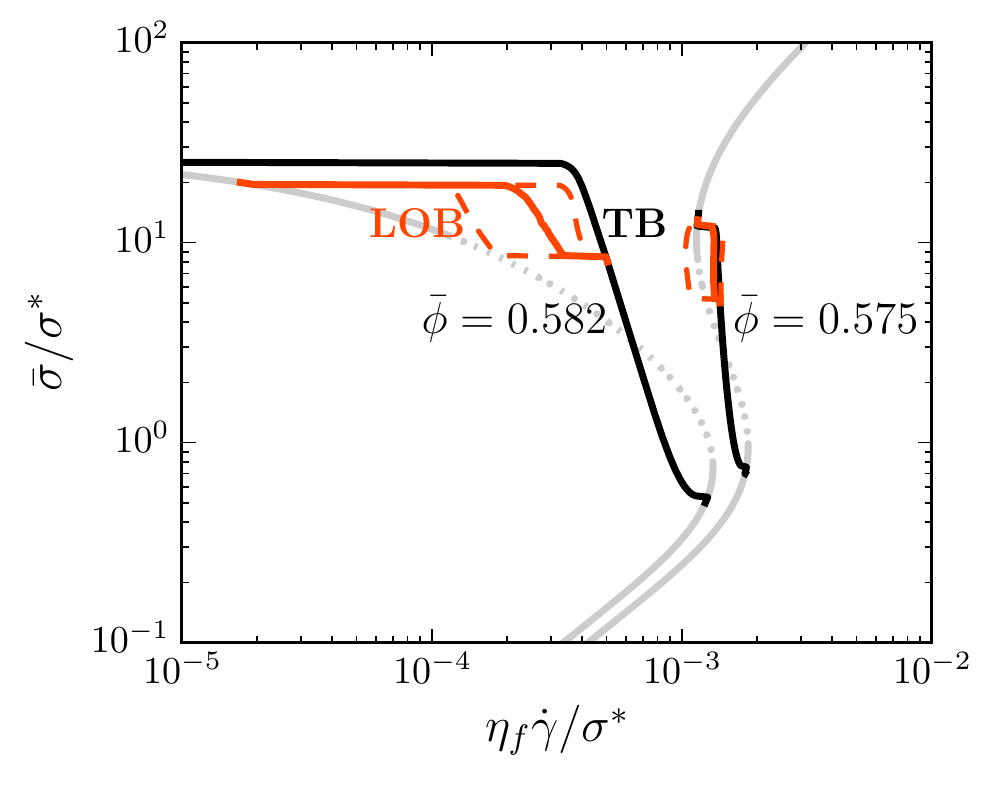}
\caption{Macroscopic flow curves for $\alpha=10^9$ with $\bar\phi=0.575$ and $\bar\phi=0.582$, obtained by sweeping stress up and down from a TB state (black lines) and from a LOB state (red lines). For LOB, solid lines are the time-averaged flow curves, while dashed lines are the low and high limits of shear rate oscillations. The underlying homogeneous steady state flow curves (grey lines) show domains of linear stability (solid lines) and instability (dashed lines).\label{meta_fig}}
\end{figure}

For the dependence of the fraction of frictional contacts $f$ on
stress $\sigma_{zz}$, particle simulations suggest a relation
\begin{equation}
\label{steady}
f^{\mathrm{SS}}(\sigma_{zz}) = \exp(-\sigma^{\ast}/\sigma_{zz})
\end{equation}
in steady state, where $\sigma^{\ast}=C F^{\ast}/a^2$. This depends
on
the typical repulsive force $F^\ast$ that must be overcome to create a
contact and the typical particle radius $a$;
particle simulations suggest $C\approx 1.45$~\cite{Mari2014,Singh2017}.
Departing now from the steady-state assumptions of ~\cite{Wyart2014},
we assume that $f$ does not react infinitely fast to
changes in stress~\cite{Mari2015a, Han2018}, following
instead
\begin{equation}
\partial_t f = - \frac{\dot{\gamma}}{\gamma_0} \left[ f - f^{\mathrm{SS}} \right] \label{dyn_f} \\.
\end{equation}
Note that this evolution involves a characteristic strain scale
$\gamma_0$, of order the strain required to evolve from one steady-state to another (for instance on flow reversal~\cite{chacko_shear_2018}). For the typical volume fractions considered here, particle
simulations suggest $\gamma_0=O(10^{-2})$~\cite{Mari2015a}.

Next we assume that particle migration between vorticity bands is driven by the
difference in normal stress $\sigma_{zz}$ that will in general
exist across the interface between them. We model this via
a ``two-fluid"~\cite{DoiBook} or ``suspension balance"
model~\cite{Nott1994,Morris1999,Yurkovetsky2008,Lhuillier2009,Nott2011}.
The divergence of the particle stress gives a force imbalance on the
particle phase, which must be rebalanced by a drag between the
particles (p) and fluid (f) due to an interphase relative velocity
$v^{\mathrm{p}}_z-v^{\mathrm{f}}_z = v^{\mathrm{p}}_z/(1-\phi)$, found 
using $\bar{v}_z =\phi v^{\mathrm{p}}_z + (1-\phi)v^{\mathrm{f}}_z$. The
resulting balance condition, $\partial_z
\sigma_{zz} = -\phi \,\alpha\, v^{\mathrm{p}}_z$, involving an interphase drag parameter $\alpha$,
then implies that particles migrate from regions of high to low stress.

Conservation of mass now imposes $\partial_t \phi + \partial_z
(v^{\mathrm{p}}_z
\phi) = 0$, which gives
\begin{equation}
\label{eq:momentum_conserv}
\partial_t \phi=  \frac{1}{\alpha}\partial_z^2\sigma_{zz}.
\end{equation}
Particle simulations~\cite{Hoef2005} suggest the drag coefficient
$\alpha$ ranges from $4.5 \eta_{\mathrm{f}} a^{-2}$ for $\phi\to 0$ to $225
\eta_{\mathrm{f}} a^{-2}$ for $\phi=0.64$.
However, in this work, variations in $\phi$ will be $5
\%$ or less, so we treat $\alpha$ as a $\phi$-independent model parameter.

Eqs.~\ref{stress}--\ref{eq:momentum_conserv} define our model.
It contains the parameters
$\eta_{\mathrm{f}},\nu,\phi_{\mathrm{J}}^\mu,\phi_{\mathrm{J}}^0,\sigma^{\ast},\gamma_0$ and $\alpha$,
along with the cell length in the vorticity direction $L_z$, the
global volume fraction $\bar{\phi} = L_z^{-1} \int_0^{L_z}
\mathrm{d}z\ \phi$, and the global mean
particle stress $\bar{\sigma} = L_z^{-1}\int_0^{L_z}
\mathrm{d}z\ \sigma_{zz}$ as imposed at the walls.
We choose $L_z$ as the length unit,
 $\sigma^*$ as the stress unit, and $\eta_{\mathrm{f}}/\sigma^*$ as the time unit.
Except when explicitly comparing with the simulation data,
we also choose to rescale all strains by $\gamma_0$,
so setting $\gamma_0=1$.
We set rheological parameters compatible with the ones
of spherical particles with moderate polydispersity, setting $\nu=2.0$~\cite{Boyer2011,Singh2017},
$\phi_{\mathrm{J}}^\mu=0.58$ (for friction $\mu\approx1$~\cite{Boyer2011,Mari2014})
and $\phi_{\mathrm{J}}^0=0.64$~\cite{scott_packing_1960,Berryman1983}.
There then remain just three dimensionless parameters: a rescaled drag
$\tilde{\alpha} = L_z^2\alpha/\eta_{\mathrm{f}}\gamma_0$
(effectively a measure of the system size $L_z/a$),
rescaled stress $\tilde{\sigma}=\bar{\sigma}/\sigma^*$ and
volume fraction $\bar{\phi}$.
We drop tildes and denote these $\alpha,\bar{\sigma}$ and $\bar{\phi}$ hereafter.

The overbars are in turn dropped when discussing strictly homogeneous, unbanded steady states, as described by the constitutive curves $\sigma(\gdot)$. These are just the stationary solutions of
Eqs.~\ref{stress}--\ref{dyn_f}, and coincide directly with those of~\cite{Wyart2014}. They  are shown as black lines in
Fig.~\ref{fc_fig}. At low volume fraction $\phi <
\phi_{\mathrm{DST}}$, they  are monotonic. For
$\phi_{\mathrm{DST}} < \phi < \phi_{\mathrm{J}}^\mu$ they are S-shaped, with a
regime in which $\mathrm{d}\sigma/\mathrm{d}\gdot<0$, giving discontinuous shear
thickening. At even larger $\phi > \phi_{\mathrm{J}}^\mu$, they bend right back
to ascend the axis $\gdot=0$ above a $\phi$-dependent shear jamming
stress $\sigma_{\mathrm{J}}$, with flow only possible for
stresses $\sigma<\sigma_{\mathrm{J}}$.

For any initial state on such a constitutive curve, the volume
fraction $\phi$ and fraction of frictional contacts $f$ are defined to be
uniform.  We now perform a linear stability
analysis to determine whether any such ``base state'' is stable, by
adding to it small-amplitude perturbations $\propto e^{i k z}$ in both $f$ and $\phi$. Expanding Eqs.~\ref{stress}--\ref{eq:momentum_conserv}
to first order in the corresponding amplitudes we find linear instability
\footnote{Linear instability arises for positive real part $\Re\, \omega > 0$ for the eigenvalue $\omega$ of the linearized dynamics. An oscillatory component arises for 
non-zero imaginary part $|\Im\, \omega| > 0$,
}
\begin{equation}
\label{eqn:unstable}
\left[\frac{\mathrm{d}{\sigma}}{\mathrm{d} \gdot}\right]^{-1} < - \frac{k^2}{\alpha} \frac{1}{\eta}\frac{\partial \eta}{\partial \phi}.
\end{equation}
In regimes of high $\sigma$ or low $\phi$, we also find an oscillatory component to the growing perturbations.
For an infinitely large system, that is, $\alpha\to \infty$,
this yields the familiar mechanical instability criterion for vorticity banding, $\mathrm{d}\sigma / \mathrm{d}\dot{\gamma} < 0$.
When the system size is finite, the
unstable region shrinks, with
stability boundaries shown as colored lines in
Fig.~\ref{fc_fig}.

The mechanism of this instability is as follows. Temporarily
ignoring variations in $\phi$, Eqs.~\ref{stress}--\ref{dyn_f}
effectively reduce to (a) $\gdot=\sigma/\eta(f)$ and (b)
$\dot{f}=- \dot \gamma [f-f^{\mathrm{SS}}(\sigma)]$. Recalling that $\gdot$ must remain uniform
in $z$ while $\sigma$ and $f$ can vary, we imagine a localised
fluctuation in which $\sigma$ slightly increases at some
$z$: (b) then requires that $f$ correspondingly also increases. For
large enough $\mathrm{d}\eta/\mathrm{d}f>0$, $\sigma$ must increase even further to maintain uniform $\gdot$ along $z$ via (a). This gives positive feedback
and instability, irrespective of wave vector $k$. We now relax the assumption of constant $\phi$,  noting that the $\phi$
relaxation depends on $k$, and is always stabilizing:
whenever $\sigma$ increases locally, there is an outward migration of
particles, thereby locally decreasing $\phi$ and hence the viscosity.
The competition between these processes restricts the unstable $f$
dynamics to large length scales, because the time for particle
migration increases with distance.

\begin{figure}
\centering
\hspace*{-3mm}
\includegraphics[width=.5\textwidth]{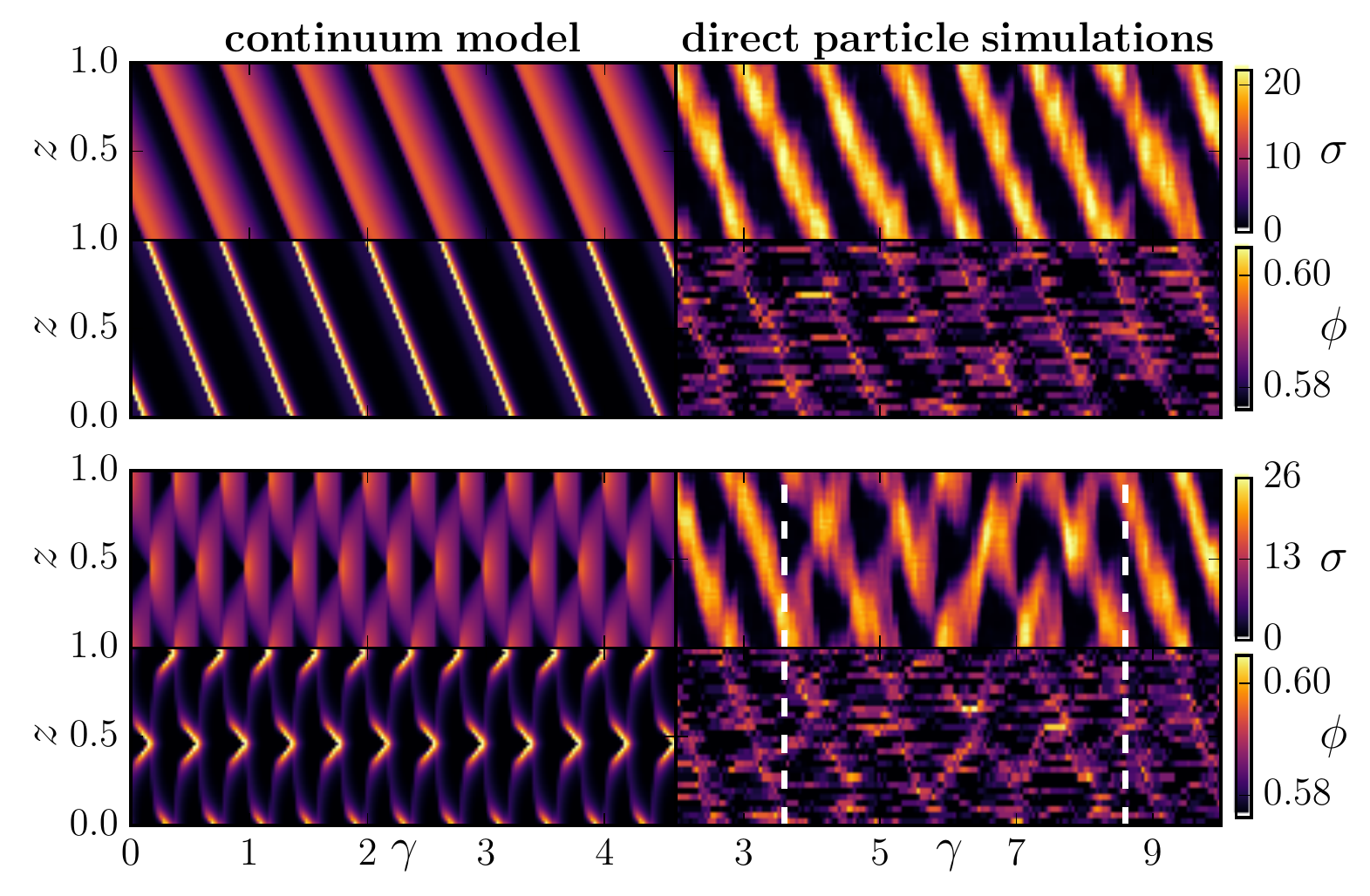}
\caption{Comparison of volume fraction and stress space-strain plots
between long-time inhomogeneous flows in the (left) continuum
model and (right) direct particle simulations at a volume
fraction $\bar\phi=0.58$. The parameters for the continuum model are $\gamma_0 = 0.023$ and $\alpha=\num{1.3e8}$,
while the particle simulation use $L_z/a=815$ and we measure $\gamma_0 = 0.023$~\cite{Mari2015a} and $\alpha = \num[separate-uncertainty]{1.4 +- 0.3 e8}$.
TB solutions at an imposed stress $\bar{\sigma}=6.525$ (top)
and LOB solutions at $\bar{\sigma}=7.25$ (bottom).
 (All stresses are nondimensionalized by setting $\sigma^* = 1$.)
LOB are only visible transiently in the simulations.
Note the different scales for the strain in the continuum model and the particle simulations:
the model predicts bands moving roughly twice faster than in the simulations.\label{fig:bands_colormaps}
}
\end{figure}

Having shown a state of initially homogeneous flow to be linearly
unstable if Eq.~(\ref{eqn:unstable}) is satisfied, we now numerically
integrate the model equations to elucidate the full nonlinear dynamics
that prevails at long times. We do so for a representative value of $\alpha=10^9$
and for two volume fractions:
$\bar\phi=0.575$, for which the constitutive curve is S-shaped; and
$\bar\phi=0.582$, for which it folds right back to the $\gdot=0$ axis.
Taking as our initial condition a state of homogeneous shear on the
constitutive curve for some imposed $\bar\sigma$, subject to small-amplitude perturbations, we find one
of two possible competing long-time states of dynamical vorticity bands:
a locally oscillating band (LOB) state for imposed shear stresses in the vicinity of
$\bar\sigma = 10$, and a travelling band (TB) state otherwise. Starting from the LOB (resp. TB) state,
we then quasi-statically sweep $\bar\sigma$ up and (in a separate run) down
from these values, generating the red (resp. black) macroscopic flow curves
shown in Fig.~\ref{meta_fig} (see~\cite{SM}
for details).

A TB state, pertaining to the black flow curve, is shown in Fig.~\ref{fig:bands_colormaps} (top left).
Here the steady-state bulk shear rate $\dot\gamma (t) = \mathrm{const.}$, and localised pulses travel along the vorticity axis at constant speed in one
direction. (The direction represents a spontaneously broken symmetry,
depending sensitively on the initial noise.) 
An LOB state, pertaining to the red flow curve, is shown in Fig.~\ref{fig:bands_colormaps} (bottom left).
Here the bulk shear rate at constant imposed stress shows sustained oscillations in time, reminiscent of experimental observations~\cite{Hermes2016,Bossis2017}.
In Fig.~\ref{meta_fig}, the average of the oscillation is shown by the solid red line,
and the limits by the dotted red lines.
These states lie within the region of oscillatory linear instability; spatiotemporally they exhibit locally oscillating bands comprising two excitations that travel in opposite
directions and intermittently collide. Such collisions coincide with a drop in the oscillating bulk shear rate
signal.
For the higher volume fraction, $\bar\phi=0.582$, homogeneous flow is recovered
above some stress threshold $\sigma = \sigma^{\rm hom} <\sigma_{\mathrm{J}}$;
but flow then arrests completely at $\sigma\ge\sigma_{\mathrm{J}}$, where the underlying constitutive curve re-joins the vertical axis \footnote{We note that the composite (macroscopic) flow
curves are tilted -- as opposed to vertical, as would have been expected
without concentration coupling~\cite{Olmsted2008} -- and that
negatively sloping macroscopic flow curves were seen experimentally in~\cite{Pan2015}.}.

We now compare these continuum results to direct particle simulations.
These use 8000 bidisperse spheres (radii $a$ and $1.4a$ in equal
volume proportions), sheared under constant  global shear stress $\bar\sigma_{xy}$ in a tri-periodic box of size
$L_x=L_y=10.2a$ and $L_z=815a$ to allow for fully developed vorticity bands~\cite{SM}, using Lees-Edwards boundary
conditions~\cite{Lees1972}.
(Note that $\bar \sigma_{zz}$ was controlled in the continuum model; this should not matter as previously discussed.)
 The spheres interact through
both lubrication and contact forces~\cite{Mari2014}, with any contact
force becoming frictional (with friction coefficient $\mu=1$) once the normal part exceeds a
fixed value $F^{\ast}$ \cite{SM}. This ``Critical Load Model''
captures both DST and jamming, but previous work~\cite{Mari2014} used
much smaller $L_z$ values, precluding the vorticity
instabilities addressed here.
With $\mu=1$, the values of $\nu$, $\phi_{\mathrm{J}}^\mu$ and $\phi_{\mathrm{J}}^0$
are consistent with the ones we set for the continuum model,
giving a good agreement for the homogeneous steady state flow curves~\cite{Singh2017}.

In our large-$L_z$ simulations, we indeed
find dynamic vorticity banding, with two distinct
dynamical states, in close analogy with those of our continuum
model. As seen in the top right of Fig.~\ref{fig:bands_colormaps}, we
recover the TB solutions at lower stresses.
We find good qualitative agreement
between simulations and our continuum model (top left of Fig.~\ref{fig:bands_colormaps})
in the overall dynamics of stress and volume fraction fields.
To perform the comparison, we used model parameters $\gamma_0 = 0.023$ and $\alpha=\num{1.3e8}$,
based on their measured value in the simulations \footnote{Specifically, we take $\gamma_0$ from~\cite{Mari2015a},
and take $\alpha = \num[separate-uncertainty]{1.4 +- 0.3 e8}$ from fits to the measured relation
between the particle phase velocity and stress gradient
$\partial_z \sigma_{zz} = -\phi \,\alpha\, v^{\mathrm{p}}_z$.}.
Both continuum model and
particle simulations show slight accumulation of particles
at the front of the travelling thickened band, albeit with somewhat flatter stress profiles in the simulations
than in the model (not shown).
The continuum model's prediction for the speed of the TB is in good qualitative agreement with our particle simulations, but roughly a factor 2 larger. 
At higher stresses, our
simulations also exhibit the LOB states of the continuum model, but we
have so far only found these as a transient effect in the particle
simulations, with LOB always eventually giving way to TB: see the
lower panel of Fig.~\ref{fig:bands_colormaps}.

In summary, we have proposed a continuum model for the vorticity instabilities of a
shear-thickening suspension held at a constant macroscopic stress
in the unstable part of the constitutive curve where $\mathrm{d}\sigma/\mathrm{d}\gdot <0$. Its
predictions compare very well with our particle based simulations, including a regime of oscillating macroscopic shear rates as
found
experimentally~\cite{Hermes2016,Bossis2017}. Crucially,
the unsteady behavior results from a bulk rheological mechanism, not from coupling with the mechanical response of the rheometer, even
if the latter plays a part in some experiments~\cite{Bossis2017}.  Particle migration is crucial: the banding dynamics relies on small
concentration variations that have large rheological effects close to jamming, as reported previously for colloidal glasses in pipe flow~\cite{Besseling2010a}.

Observing the predicted spatiotemporal bands directly in experiments, by measuring local stress fields and small concentration fluctuations, may prove challenging.
The velocity field along the vorticity direction also bears a signature of the bands
due to particle migration, which could be more accessible.
Very recent experiments do report vorticity bands in cornstarch suspensions under controlled stress,
similar to those presented here in shape, size and velocity~\cite{Saint-Michel2018}.
However, the banding signature involves the flow ($v_x$) velocity component; this
may stem from differential wall slip induced
by a frictional band moving along the vorticity direction.

While we focused here on a constitutive model involving only the normal stress along the vorticity direction,
the physical ingredients in our model may also admit instabilities
along the gradient and/or flow directions. Without volume fraction
variations, however, homogeneous shear flow is predicted to be stable
against gradient perturbations in the absence of inertia [18], for
constitutive curves of the shapes considered here. Any instability in
the gradient direction would therefore have to be driven by particle
migration (and/or inertial) effects, and is likely to be subdominant
to the vorticity banding considered here.
 Indeed some experiments in the discontinuous shear thickening regime report
a flowing gradient band of lower concentration coexisting with a densely jammed band~\cite{Fall2015}.
For very large systems, inertia will separately trigger gradient instabilities~\cite{Nakanishi2012}.
Our coupled model of shear thickening and particle migration represents a first step towards explaining the full range
of unsteady flows close to the jamming transition in dense suspensions and, we hope, will stimulate systematic experimental studies of this regime.

We thank M. Hermes, I. Peters, B. Saint-Michel, T. Gibaud and
S. Manneville for insightful discussions.  Work funded in part by SOFI CDT, Durham University,
and EPSRC (EP/L015536/1); the European
Research Council under the European Union's Seventh Framework
Programme (FP7/2007-2013) and Horizon 2020 Programme /  ERC grant agreements 279365 and 740269.  MEC
is funded by the Royal Society.

\bibliographystyle{apsrev4-1}

\end{document}